\documentclass{article}
\usepackage{amsmath}
\usepackage[psamsfonts]{amssymb}
\usepackage{cmmib57}
\usepackage[cmtip,arrow]{xy}
\usepackage{pb-diagram,pb-xy}
\newcommand{\bPf}{\par\vspace*{-4pt}\indent{\sc Proof.}\enskip}
\newcommand{\ePf}{\medskip}
\def\QED{\hskip0.1em\hfill\null\ \null\nobreak\hfill\kern3pt\vbox{\hrule\hbox
   {\vrule\kern1pt\vbox{\kern1.7pt\hbox{$\scriptscriptstyle{QED}$}
    \kern0.2pt}\kern1pt\vrule}\hrule}}

\def\END{\hskip0.1em\hfill\null\ \null\nobreak\hfill\kern3pt\vbox{\hrule\hbox
   {\vrule\kern1pt\vbox{\kern1.7pt\hbox{$\,\,\,\vspace{5pt}$}
    \kern0.2pt}\kern1pt\vrule}\hrule}}
\newtheorem{theorem}{Theorem}
\newtheorem{lemma}{Lemma}
\newtheorem{corollary}{Corollary}
\newtheorem{proposition}{Proposition}
\newtheorem{remark}{Remark}
\newtheorem{definition}{Definition}
\newtheorem{example}{Example}
\newcommand{\bCd}{\bEq\begin{CD}}
\newcommand{\eCd}{\end{CD}\eEq}
\newcommand{\bcd}{\beq\begin{CD}}
\newcommand{\ecd}{\end{CD}\eeq}
\newcommand{\ben}{\begin{enumerate}}
\newcommand{\een}{\end{enumerate}}
\newcommand{\bEq}{\begin{eqnarray}}
\newcommand{\eEq}{\end{eqnarray}}
\newcommand{\beq}{\begin{eqnarray*}}
\newcommand{\eeq}{\end{eqnarray*}}
\newcommand{\bDf}{\begin{definition}\em}
\newcommand{\eDf}{\end{definition}}
\newcommand{\bLm}{\begin{lemma}}
\newcommand{\eLm}{\end{lemma}}
\newcommand{\bPr}{\begin{proposition}}
\newcommand{\ePr}{\end{proposition}}
\newcommand{\bTh}{\begin{theorem}}
\newcommand{\eTh}{\end{theorem}}
\newcommand{\bCr}{\begin{corollary}}
\newcommand{\eCr}{\end{corollary}}
\newcommand{\bRm}{\begin{remark}\em}
\newcommand{\eRm}{\end{remark}}
\newcommand{\bEx}{\begin{example}\em}
\newcommand{\eEx}{\end{example}}

\newcommand{\Z}{\mathbb{Z}}
\newcommand{\ie}{{\em i.e$.$} }
\newcommand{\eg}{{\em e.g$.$} }
\newcommand{\R}{I\!\!R}

\newcommand{\cE}{\mathcal{E}}

\newcommand{\cL}{\mathcal{L}}

\newcommand{\bK}{\boldsymbol{K}}

\newcommand{\bU}{\boldsymbol{U}}

\newcommand{\bX}{\boldsymbol{X}}
\newcommand{\bY}{\boldsymbol{Y}}

\newcommand{\sub}{\subset}

\newcommand{\wed}{\wedge}
\newcommand{\com}{\!\circ\!}

\newcommand{\bet}{\beta}
\newcommand{\gam}{\gamma}
\newcommand{\del}{\delta}
\newcommand{\eps}{\epsilon}

\newcommand{\lam}{\lambda}

\newcommand{\ome}{\omega}

\newcommand{\For}{{\Lambda}}
\newcommand{\Con}{{\mathcal{C}}}
\newcommand{\Hor}{{\mathcal{H}}}
\newcommand{\Var}{{\mathcal{V}}}
\newcommand{\Thd}{{\Theta}}
\title{{\Large {\bf Variationally equivalent problems and variations of Noether currents}}}

\author{M. Francaviglia, M. Palese  and  E. Winterroth \\
{\footnotesize Department of Mathematics, University of Torino} \\ {\footnotesize via C. Alberto 10, I-10123 Torino, Italy } \\  {\footnotesize {\sc e-mail: [marcella.palese, ekkehart.winterroth]@unito.it}}}

\date{}

\begin{document}

\maketitle

\begin{abstract}
We consider systems of local
variational problems defining non vanishing cohomolgy classes.
In particular, we prove that the conserved current associated with a generalized symmetry, assumed to be also a symmetry of the variation of the corresponding local inverse problem, is variationally equivalent to the variation of the strong Noether current for the corresponding local system of Lagrangians. This current is conserved and a sufficient condition will be identified in order such a current be global.

\medskip



\end{abstract}

%-----------------------------------------------------------------------------%
\section{Introduction}
%-----------------------------------------------------------------------------%

The aim of this paper is the study of Noether conservation laws associated with  some `variational' invariance of global Euler-Lagrange morphisms of local problems of a given type.
In this context, the question arises whether we would be interested in conservation laws different from those directly associated with invariance properties of field equations. 
The answer to this question relays on Emmy Noether's famous and celebrated  paper {\em Invariante Variationsprobleme} \cite{Noe18}.
As it is well known, Noether's paper was motivated by the fact that, although the gravitational field equations were global, the associated conservation laws found by Einstein by a nonvariational approach were not (think of the well known energy-momentum {\em pseudo-tensor}).
The underlying idea was that of looking for conservation laws coming from invariance properties of a ({\em  possibly local}) Lagrangian (rather than a field equation solely) to find a way of associating global conservation laws with the gravitational {\em field}.

Explicitly, in the introduction of her paper, Noether  wrote: {\em \"Uber diese aus Variationsproblemen entspringenden Differentialgleichungen lassen sich viel pr\"azisere Aussagen machen als \"uber beliebige, eine Gruppe gestattende Differentialgleichungen, die den Gegenstand der Lieschen Untersuchungen bilden}; so underlining the relevance of the study of differential equations generated by an invariant  variational problem in its whole rather than of arbitrary differential equations admitting a Lie group of symmetries. It is pointed out that considering invariance of variational problems the issue is a major refinement in the results: to symmetries of equations could correspond conservation laws which have a nonvariational meaning and thus could not be characterized in a similar  precise manner.

In fact, contrarily to what sometime improperly stated, Noether's Theorem (I and II Theorems actually) were not `formulated for Euler-Lagrange equations in field theory', rather they are statements about the invariance of the variational problem (in other words, of the action integral) with respect to a finite continuous group of transformations and an infinite continuous group of transformations, respectively. 
The direct object of Noether's investigations are what she calls {\em Lagrangeschen Ausdr\"ucke; d.h. die linken Seiten der Lagrangeschen Gleichungen}, which we shall call hereafter Euler-Lagrange expressions. The accent is not put on field {\em equations} although her results have, of course, {\em also} consequences concerning invariance properties of equations. It is maybe noteworthy that all Noether considerations are made `off shell', \ie not along solutions of Euler-Lagrange equations; in particular it is an evident matter that, `on shell' Noether identities would reduce to the trivial identity $0=0$. It is also important to stress that Noether immediately consider the formulation of a variational problem at an infinitesimal level ({\em integralfreie Identit\"at}).

Noether's Theorem II is in fact concerned with a variation of the Euler-Lagrange expressions. Symmetry properties of the Euler-Lagrange expressions play a fundamental role since they  introduce {\em a cohomology class which adds up to Noether currents}\,\footnote{A formulation in modern language of Noether's results.}; they are related with invariance properties of the first variation, thus with the vanishing of a second variational derivative. As we shall see later on, the concept of a variation of Noether current is then clearly involved.
In this paper, therefore, symmetries of Euler-Lagrange morphisms or, more generally, of so-called variational `dynamical forms' are considered insomuch as they can provide informations about Noether currents of some potential Lagrangians, also in the spirit of the Bessel-Hagen version of Noether's Theorem II \cite{BeHa21}. We would like to stress that a Noether current is not necessarily a {\em conserved current}. We shall investigate under which conditions the variation of a strong Noether current is conserved: the vanishing of the second variational derivative will be involved; but an additional sufficient condition will be identified in order such a current be global.

Geometric definitions of conserved quantities
in field theories have been proposed within formulations based on symmetries of Euler-Lagrange operator rather than of the Lagrangian (see \eg \cite{Sar03,Sar03b}). 
In particular, a definition of {\em variation of conserved currents} for  gauge-natural theories \cite{Eck81} has been proposed in \cite{FeFrRa03} where the first variation of a `deformed' (also called there `variational') Lagrangian -- obtained by contracting the Euler-Lagrange term with the infinitesimal symmetry defined by preferred lifts of infinitesimal principal automorphisms (gauge-natural lifts) -- was considered. 
On the other hand, without the fixing of a connection {\em a priori}, the gauge-natural second variational derivative of gauge-natural invariant Lagrangian has been exploited in order to construct Noether covariant conserved current \cite{PaWi03, PaWi04}, by showing the relationship of the Bergmann--Bianchi morphism with the variational derivative of a conveniently deformed (gauge-natural) Lagrangian and with Noether identities: generalized symmetries and some version of Noether's Theorem II play a fundamental role \cite{FFPW08,FFPW11}; see also \cite{Sar09} for a characterization of superpotentials in gauge theories. Strictly related with the present approach are also papers proposing the concept of relative conservation laws; see \eg \cite{FaFeFr05}.
 
Inspired by Noether's results and by Lepage's seminal paper \cite{Lep36}, geometrical formulations of the calculus of variations
on fibered manifolds constitute a large class of theories
for which the Euler--Lagrange operator is a morphism of an exact
sequence, see \eg \cite{AnDu80,DeTu80,Kru90,Tak79,Tul77,Vin77}. The module in degree $(n+1)$, contains so-called  (variational) dynamical forms; a given equation is globally an Euler--Lagrange equation if its dynamical form is the differential of a Lagrangian and this is equivalent to the dynamical form being closed in the complex, \ie Helmholtz conditions hold true, and its cohomolgy class being trivial.
Dynamical forms which are only {\em locally variational}, \ie which are closed in the complex and define a non trivial cohomology class, admit a system of local Lagrangians, one for each open set in a suitable covering, which satisfy certain relations among them.
Global projectable vector field on a jet fiber manifold which are symmetries of dynamical forms, in particular of locally variational dynamical forms, and corresponding formulations of Noether theorem II can be considered  in order to determine obstructions to the globality of associated conserved quantities \cite{FePaWi10}. Analogously, the concept of global (and local) variationally trivial Lagrangians and, in general, of variationally trivial currents (\ie $(n-1)$-forms) will be fundamental in this paper. 

In fact, in the sequel, a {\em locally variational form} is any closed $p$-form in the variational sequence; we will be faced with inverse problems at any degree of variational forms.

The role, in the cohomology, of the {\em variational Lie derivative}, a differential operator acting on equivalence classes of forms in the variational sequence, has very relevant implications in the calculus of variations.
Iterated variational derivatives define higher order variations \cite{FrPaWi05,PaWi03,PaWi07}; thus variations of currents can be recognized in this approach.
Consequently, we shall now study variations of conserved currents in a quite general setting. 

%-----------------------------------------------------------------------------%
\section{Equivalence of local variational problems}
%-----------------------------------------------------------------------------%

We shall consider the variational sequence \cite{Kru90} defined on a fibered manifold $\pi: \bY \to \bX$, with $\dim
\bX = n$ and $\dim \bY = n+m$. For $r \geq 0$ we have
the $r$--jet space $J_r\bY$ of jet prolongations of sections of 
the fibered manifold $\pi$. We have also the natural fiberings $\pi^r_s : J_r\bY \to
J_s\bY$, $r \geq s$, and $\pi^r : J_r\bY \to \bX$; among these the
fiberings $\pi^r_{r-1}$ are {\em affine bundles} which  induce the natural fibered splitting 
\beq 
J_r\bY\times_{J_{r-1}\bY}T^*J_{r-1}\bY \simeq
J_r\bY\times_{J_{r-1}\bY}(T^*\bX \oplus V^*J_{r-1}\bY)\,,
\eeq
which, in turn,  induces also a decomposition of the exterior
differential on $\bY$ in the {\em horizontal\/} and {\em vertical
differential\/}, $(\pi^{r+1}_r)^*\com\, d = d_H + d_V$.
By $(j_{r}\Xi, \xi)$ we denote the jet
prolongation of a {\em projectable vector field} $(\Xi, \xi)$ on $\bY$, and by $j_{r}\Xi_{H}$ and
$j_{r}\Xi_{V}$ the horizontal and the vertical part
of $j_{r}\Xi$, respectively.

We have  the {\em sheaf splitting}
$\Hor^{p}_{(s+1,s)}$ $=$ $\bigoplus_{t=0}^p$
$\Con^{p-t}_{(s+1,s)}$ $\wed\Hor^{t}_{s+1}$ where 
$\Hor^{p}_{(s,q)}$ and
$\Hor^{p}_{s}$  ($q \leq s $) are sheaves of {\em horizontal forms},  and 
 $\Con^{p}_{(s,q)}
\sub \Hor^{p}_{(s,q)}$ are subsheaves of {\em contact forms} \cite{Kru90}. Let us denote by $h$ the projection onto the nontrivial summand with the higest value of $t$ and  by $d\ker h$ the sheaf generated by
the corresponding presheaf and set then $\Thd^{*}_{r}$ $\equiv$ $\ker h$ $+$
$d\ker h$;
the quotient sequence
\beq
0\arrow{e} \R_{\bY} \arrow{e} \dots\,\
\arrow[4]{e,t}{\cE_{n-1}}\,\ \ \For^{n}_r/\Thd^{n}_r
\arrow[3]{e,t}{\cE_{n}}\,\ \For^{n+1}_r/\Thd^{n+1}_r
\arrow[4]{e,t}{\cE_{n+1}}\,\ \ \For^{n+2}_r/\Thd^{n+2}_r
\arrow[4]{e,t}{\cE_{n+2}}\,\ \ \dots\,\ \arrow{e,t}{d} 0
\eeq
defines the {\em $r$--th order variational sequence\/}
associated with the fibered manifold $\bY\to\bX$; here $\For^{p}_{s}$
 is obviously the standard sheaf of $p$--forms on $J_s\bY$. The quotient sheaves (the sections of which are classes of forms modulo contact forms) in the variational sequence can be represented as sheaves $\Var^{k}_{r}$ of $k$-forms on jet spaces of higher order. In particular, currents are classes  $\nu\in(\Var^{n-1}_{r})_{\bY}$;
Lagrangians are classes $\lam\in(\Var^{n}_{r})_{\bY}$, while $\cE_n(\lam)$ is called a Euler--Lagrange form (being 
$\cE_{n}$ the Euler--Lagrange morphism); dynamical forms are classes $\eta\in(\Var^{n+1}_{r})_{\bY}$ and $\cE_{n+1}(\eta)$ is a Helmohltz form (being $\cE_{n+1}$ the corresponding Helmholtz morphism).

It turns out that the variational sequence is an exact resolution of the constant sheaf $\R_{\bY}$
over $\bY$.
Since the variational sequence is a soft
resolution of the constant sheaf $\R_{\bY}$ over $\bY$,  the cohomology of the complex of global sections, denoted by $H^*_{VS}(\bY)$,  is naturally isomorphic to both the \v Cech cohomology of  $\bY$ with coefficients in the constant sheaf $\R$ 
and  the de Rham cohomology $ H^*_{dR}\bY$ \cite{Kru90}.

Let $\bK^{p}_{r}\doteq \textstyle{Ker}\,\, \cE_{p}$.  We have the short exact sequence of sheaves
\beq 
0 \arrow{e}\bK^{p}_{r}\arrow[2]{e,t}{i}\, \, \Var^{p}_{r}\arrow[2]{e,t}{\cE_{p}}\, \
\cE_{p}(\Var^{p}_{r})\arrow{e} 0 \,.
\eeq 

For any global section $\beta\in(\Var ^{p+1}_{r})_{\bY}$ we have 
$\beta\in(\cE_{p}(\Var^{p}_{r}))_{\bY}$ if and only if $\cE_{p+1}(\beta)=0$, which are conditions of local variationality.
A global inverse problem is to find necessary and sufficient conditions for such a locally variational $\beta$ to be globally variational.
In particular, $\cE_{n}(\Var^{n}_{r})$ is the sheave of Euler--Lagrange morphisms and $\eta\in(\cE_{n}(\Var^{n}_{r}))_{\bY}$ if and only if $\cE_{n+1}(\eta)=0$, which are Helmholtz conditions.

The above exact sequence gives rise to the long exact
sequence in \v Cech cohomology 
\beq 
0 \arrow{e} (\bK^{p}_{r})_{\bY} \arrow{e}
(\Var^{p}_{r})_{\bY} \arrow{e}
(\cE_{p}(\Var^{p}_{r}))_{\bY} \arrow{e,t}{\del_{p}} H^{1}(\bY,
\bK^{p}_{r})\arrow{e} 0 \,. 
\eeq 

In particular, every $\eta\in(\cE_{n}(\Var^{n}_{r}))_{\bY}$ (\ie locally variational) defines a cohomology class 
$\del (\eta) \equiv  \del_{n} (\eta) \in
H^{1}(\bY, \bK^{n}_{r}) $ $\simeq$ $ H^{n+1}_{VS}(\bY)
$ $\simeq$ $ H^{n+1}(\bY,\R)$.
Furthermore, every $\mu\in(d_H(\Var^{n-1}_{r}))_{\bY}$ (\ie locally variationally trivial) defines a cohomology class 
$ \del' (\mu) \equiv  \del_{n-1} (\mu)\in
H^{1}(\bY,  \bK^{n-1}_{r}) $ $\simeq$ $ H^{n}_{VS}(\bY)
$ $\simeq$ $ H^{n}(\bY,\R)$.

For any  countable
open covering $\mathfrak{U}=\{U_{i}\}_{i\in I}$, $I\sub \Z$,  $C^{q}(\mathfrak{U},\Var^{p}_{r} )$ is the set of $q$--cochains with
coefficients in the sheaf $\Var^{p}_{r}$, if $\mathfrak{d}: C^{q}(\mathfrak{U},\Var^{p}_{r} )\to
C^{q+1}(\mathfrak{U},\Var^{p}_{r} )$ is the {\em coboundary operator}, the {\em connecting homomorphism} $\delta_{p} = i^{-1}\circ\mathfrak{d}\circ\mathcal{E}_{p}^{-1}$ is the mapping of cohomologies in the corresponding diagram of cochain complexes.

Note that $\eta$ is globally variational if and only if $\del (\eta) = 0$.
In the following we will be interested in the non trivial case $\del (\eta) \neq 0$ whereby $\eta =  \cE_n(\lam)$ can be solved only locally, \ie for any countable good covering of $\bY$ there exists a local Lagrangian $\lam_{i}$ over each subset $U_{i}\sub\bY$ such that $\eta_{i}=\cE_{n}(\lam_{i})$.

A {\em local variational problem} is a system of local sections  $\lam_{i}$ of  $(\Var^{n}_{r})_{U_{i}}$ such that  $\cE_{n}((\lam_{i}-\lam_{j})|_{U_{i}\cap U_{j}}) = 0$. We notice that every cohomology class in  $H^{n+1}(\bY,\R)$ gives  rise to local variational problems.
We are then faced with the situation that
$\lam= \{\lam_{i}\}_{i\in I}$ is a $0$--cochain of Lagrangians in \v Cech
cohomology with values in the sheaf $\Var^{n}_{r}$, \ie $\lam\in
C^{0}(\mathfrak{U},\Var^{n}_{r})$; we shall
denote by $\eta_{\lam}$ the $0$--cochain formed by the restrictions
$\eta_{i}=\cE_{n}(\lam_{i})$ (and so will do at any degree of forms).
Of course, if $\eta \in
C^{0}(\mathfrak{U},\Var^{n+1}_{r})$, then $\mathfrak{d}\eta =0$ if
and only if $\eta$ is global. Note that
$\mathfrak{d}\lam=0$ implies $\mathfrak{d}\eta_{\lam}=0$, while we only have $\mathfrak{d}\eta_{\lam}=\eta_{\mathfrak{d}\lam}=0$ \ie
$\mathfrak{d}\lam\in C^{1}(\mathfrak{U},\bK^{n}_r)$ \cite{BFFP03}. 

\bRm
Two local variational problems of degree $p$ are {\em equivalent } if and only if they give rise to the same variational class of forms as the image of the corresponding morphism $\cE_p$ in the variational sequence.\END\eRm

Noether's Theorems relate  symmetries of a variational problem to conserved quantities: the concept of a {\em variational Lie derivative} operator $\cL_{j_{r}\Xi}$ which is a local differential operator enables us to define symmetries  of Lagrangian and dynamical forms (as well as of higher degree classes of forms in the variational sequence) and corresponding Noether's Theorems \cite{FPV02}. We notice that the variational Lie derivative sends a diagram of cochain complexes into a diagram of cochain complexes and thus acts on cohomology classes.
The cohomology class defined by a system of local Lagrangian is related with the cohomology class defined by the local variational problem given by the system of their local variational Lie derivative: closed variational forms defining nontrivial cohomology classes are trasformed in variational forms with trivial cohomology classes  \cite{PaWi11,PaWiGa12}.

We call $(\{\bU_{i}\}_{i\in \Z}, \lam_{i})$ a {\em presentation} of the local variational problem and we remark that an infinitesimal symmetry of a local presentation is not necessarily a symmetry of another local presentation \cite{FePaWi10}.

Let us now first summarize some preliminary results, partially stated in  \cite{PaWi11,PaWiGa12}.

\bLm
Let  $\mu\in\Var^{p}_{r}$, with $p\leq n$, be a locally  variationally trivial $p$-form, \ie such that $\cE_p(\mu )= 0$ and let $\del_p(\mu_\nu) \neq 0$. We have $\del_p (\cL_{\Xi}\mu_{\nu}) = 0$.
Analogously,  let  $\eta\in\Var^{p}_{r}$, with $p\geq n+1$, be a locally variational $p$-form, \ie such that $\cE_{p}(\eta )= 0$ and let $\del_{p}(\eta_\lam) \neq 0$. We have $\del_p (\cL_{\Xi}\eta_{\lam}) = 0$.
\eLm

\bPf
In fact, we have a $0$-cocycle of currents $\nu_i$ ($\mathfrak{d}\nu_i\neq0$) such that $ \mu =d_H \nu_i$  and  $\mathfrak{d}\mu_\nu =0$;  by using the representation of the Lie derivative of classes of variational forms of degree $p\leq n-1$ given in \cite{FPV02}, we have 
\beq
\mu_{\cL_{\Xi}\nu_{i}}=d_{H} (\Xi_H \rfloor d_{H}\nu_{i}+ \Xi_V\rfloor d_{V}\nu_{i})
\eeq
on the other hand, resorting to the naturality of the variational Lie derivative, we have 
$\mu_{\cL_{\Xi}\nu_{i}} =  \cL_{\Xi}\mu_\nu $, so that $\del_p (\cL_{\Xi}\mu_{\nu}) = 0$.
\ePf

Analogously, we have a $0$-cocycle of Lagrangians (case $p=n+1$) -- or of variational forms of higher degree (in case $p=n+2$ we have a $0$-cocycle of dynamical forms) -- $\lam_{i}$ ($\mathfrak{d}\lam_i\neq0$) such that 
$\eta = \cE_p( \lam_i)$; by linearity 
\beq
\eta_{\cL_{\Xi}\lam_{i}} 
=  \cE_{n} (\Xi_V \rfloor \eta_{\lam}) \,,
\eeq
and again resorting to the naturality of the variational Lie derivative we have
$\eta_{\cL_{\Xi}\lam_{i}} = \cL_{\Xi}\eta_{\lam}$, 
so that $\del_p (\cL_{\Xi}\eta_{\lam}) = 0$.
\ePf

\bRm \label{Remark}
In particular, since we also have $\cL_{\Xi}\mu_\nu = \mu_{\cL_{\Xi}\nu_{i}}=d_H (\Xi_H \rfloor \mu_{\nu}+ \Xi_V\rfloor p_{d_{V}\mu_{\nu}}) $, from the definition of an equivalent variational problem, we can state that the local problem defined by $\cL_{\Xi}\nu_{i}$ is variationally equivalent to the global problem defined by $\Xi_H \rfloor \mu_{\nu} + \Xi_V\rfloor p_{d_{V}\mu_{\nu}}$. 
The class $\mu_{\nu}=d_H \nu_i$ is assumed to satisfy $\mathfrak{d}\mu_{\nu}=0$, \ie it is a global object. We also notice that $\Xi_V\rfloor d_{V}\nu_{i}= \Xi_V\rfloor p_{d_{V}\mu_{\nu}} +d_H \phi_i$ thus $\Xi_V\rfloor d_{V}\nu_{i}-d_H \phi_i$ is a global object and, in general,  its cohomology class can be non trivial.

Analogously, as a consequence of the fact that $\eta_{\cL_{\Xi}\lam_{i}} =\cE_{n} (\Xi_V \rfloor \eta_{\lam})$ we have that the local problem defined by the local presentation $\cL_{\Xi}\lam_{i}$ is variationally equivalent to the global problem defined by $\Xi_V \rfloor \eta_{\lam}$.
\END\eRm

%----------------------------------------------------------------------------------------------
\subsection{Variation vector fields which are generalized symmetries}
%----------------------------------------------------------------------------------------------

Let  $\eta_{\lam}$ be the global Euler--Lagrange morphism of a local variational problem.
In the sequel, we will assume $\Xi$ to be a generalized symmetry, \ie a symmetry of a class of $(n+1)$-forms $\eta$ in the variational sequence.

It is a well known fact that $\Xi$ being a generalized symmetry implies that $\cE_n (\Xi_V\rfloor\eta) = 0$, thus locally $\Xi_V\rfloor\eta=d_H\nu_i$,  then there exists a $0$-cocycle $\nu_i$,  defined by $\mu_\nu = \Xi_V\rfloor \eta_{\lam}\equiv d_H \nu_i$. 
Notice that $\mathfrak{d}(\Xi_V\rfloor \eta_{\lam})=0$, but in general $\del_n(\Xi_V\rfloor \eta_{\lam})\neq0$ \cite{FePaWi10}. 
Along critical sections this implies the conservation law $d_H \nu_i =0$. 

On the other hand Noether's Theorem II  implies that locally $\cL_{\Xi} \lam_{i}=d_H \zeta_{i}$, thus we can write
$ \Xi_{V} \rfloor \eta_{\lam}  + d_{H}( \eps_i  -  \zeta_{i} )$  $=$ $0$, 
where $\eps_i = j_{r}\Xi_{V} \rfloor p_{d_{V}\lam_{i}}+ \xi \rfloor \lam_{i}$ is the usual {\em canonical} Noether current; the current  $\eps_{i} - \zeta_{i}$ is a {\em local} object and it is conserved along the solutions of Euler--Lagrange equations (critical sections).  Note that the Noether current $\eps_{i}$ is conserved along critical sections
if and only if $\Xi$ is also a symmetry of $\lam_{i}$. It is clear that, without additional information, to determine the current $\nu_i$ is not a simple matter, specifically when $\Xi$ {\em is not} also a symmetry of the Lagrangian. 
We also stress that when $\Xi$ is only a symmetry of a dynamical form and not a symmetry of the Lagrangian, the current $\nu_i+\eps_i$ {\em is not a conserved current} and it is such that $d_H (\nu_i+\eps_i)$ is locally equal to $d_H\zeta_i$; see also \cite{Sar03}. We shall call $(\nu_i+\eps_i)$ a {\em strong Noether current}.
Notice that if $\Xi$ would be also a symmetry of the cochain of Lagrangians a strong Noether current would turn out to be a {\em conserved current} along any sections, not only along critical sections.

We can associate a system of global currents with a system of local conserved  currents by taking the Lie derivatives of the local system, for which $\del '(\cL_{\Xi}\mu_{\nu})= 
\del '(\cL_{\Xi}(\Xi_V\rfloor \eta_{\lam})) = \del '(\cL_{\Xi}(d_H\nu_i))  =  \del '(d_H(\cL_{\Xi}\nu_i))=0$ holds true. As a result we have that divergence expressions of the local problem defined by $\cL_{\Xi}\nu_{i}$ coincide with divergence expressions for the global current $\Xi_H\rfloor\Xi_V\rfloor \eta_\lam + \Xi_V \rfloor p_{d_{V}(\Xi_V\rfloor\eta_\lam)}$. 

We shall now study variations of conserved currents in a quite general setting by determining the condition for the variation of a system of local strong Noether current to be equivalent to a  system of  {\em global conserved currents}. 

\bLm
Suppose that $\mathfrak{d}\lam_i = d_H\gam_{ij}$.
The condition $\cL_\Xi \cL_\Xi \lam_i =0$ implies 
$ \cL_\Xi d_H\gam_{ij}=\mathfrak{d}d_H\bet_i$, with $\bet_i= \nu_i+\eps_i +d_H \ome_i$.
\eLm

\bPf
Since, by linearity, we have
$\cL_\Xi \mathfrak{d}\cL_\Xi \lam_i =\mathfrak{d}\cL_\Xi \cL_\Xi \lam_{i} $,
the condition 
$\cL_\Xi \cL_\Xi \lam_i =0$ implies 
$\cL_\Xi \cL_\Xi \mathfrak{d}\lam_i =0$, \ie by Noether's Theorem II  we must have $\cL_\Xi \mathfrak{d}\lam_i =d_H\zeta_{ij}$, where $\zeta_{ij}$ is the sum of the Noether current 
associated with  $\mathfrak{d}\lam_i $ and a  form locally given as $\mathfrak{d}\nu_i+d_H\rho_{ij}$.  
On the other hand 
$\cL_\Xi \mathfrak{d}\lam_i = 
\Xi_V\rfloor\cE_n (\mathfrak{d}\lam_i )+ d_H (j_{r+1}\Xi_{V}\rfloor p_{d_{V} \mathfrak{d}\lam_{i} } + \xi\rfloor \mathfrak{d}\lam_i )\equiv\mathfrak{d} d_{H} \eps_{i}$, where 
$\eps_i$ is the Noether current associated with $\lam_i $. 

Since we are assuming $\mathfrak{d}\lam_i = d_H\gam_{ij}$, we also have $\cL_\Xi \mathfrak{d}\lam_i = \cL_\Xi d_H\gam_{ij}$. 
On the other hand, since we assume $\cL_\Xi \cL_\Xi \lam_i =0$ we also have by Noether's Theorem  II $\cL_\Xi \lam_i = d_H\bet_i$, hence 
from $\mathfrak{d}\cL_\Xi \lam_i=\cL_\Xi \mathfrak{d} \lam_i$ we get, locally, $\mathfrak{d}d_H\bet_i=\cL_\Xi d_H \gam_{ij}$, \ie
(by Noether's Theorem II  we actually  have $\bet_i= \nu_i+\eps_i +d_H \ome_i$)
$\mathfrak{d}d_H(\nu_i+\eps_i)= \cL_\Xi d_H \gam_{ij}$.
Thus 
$\cL_\Xi \mathfrak{d}\lam_i =d_H\zeta_{ij}= \cL_\Xi d_H\gam_{ij}=\mathfrak{d}d_H\bet_i$.
\ePf

\bRm
The above is also equivalent to the equality
$\mathfrak{d} \eps_i +\mathfrak{d}\nu_i+ d_H\rho_{ij}= \mathfrak{d} (\nu_i+\eps_i +d_H \ome_i)$, \ie $d_H\rho_{ij}= d_H \mathfrak{d} \ome_i$.
Notice that if $d_H\zeta_{ij}=0$, \ie if $\Xi$ is a symmetry of $\mathfrak{d}\lam_i$, we have 
$\cL_\Xi d_H\gam_{ij}=0$, then 
$d_H \mathfrak{d}\bet_i=0$, which means $\mathfrak{d}\bet_i=d_H\psi_{ij}$; thus finally we can write
$\mathfrak{d}(\nu_i +\eps_i )= d_H(\psi_{ij} -\mathfrak{d}\ome_i)=d_H(\psi_{ij} -\rho_{ij})$.\END
\eRm

\bLm
Let $\Xi$ be such that $\cL_{\Xi}\eta_{\lam_i} = 0$.
If $\cL_\Xi \cL_\Xi \lam_i = 0$ and $\cL_{\Xi}\mathfrak{d}\lam_i = 0$, then we have the conservation law  $d_H \cL_\Xi  (\nu_i+\eps_i)=0$, where $\cL_\Xi  (\nu_i+\eps_i)$ is a local representative of a global conserved  current.
\eLm
\bPf
In fact, since we are supposing $\Xi$ being a {\em generalized symmetry}, we have
\beq
\cL_\Xi \cL_\Xi \lam_i = d_H \cL_\Xi  (\nu_i+\eps_i)\,.
\eeq
On the other hand $\cL_{\Xi}\mathfrak{d}\lam_i = 0$ implies $\mathfrak{d} d_H(\nu_i+\eps_i)=0$, so that the local problem defined by $\cL_\Xi  (\nu_i+\eps_i)$ is variationally equivalent to the problem defined by the the global current
$\Xi_H\rfloor \mu_{\nu+\eps} + \Xi_V \rfloor p_{d_{V}\mu_{\nu+\eps}}  \equiv
\Xi_H \rfloor d_H (\nu_i+\eps_i) + \Xi_V\rfloor p_{d_{V}(d_H (\nu_i+\eps_i) )} \,.
$
\ePf

\bRm
Notice that the condition $\cL_\Xi \cL_\Xi \lam_i = 0$ means that the symmetry $\Xi$ is not only  a generalized symmetry, but it is required to be {\em also a symmetry of the local variational problem $\cL_\Xi \lam_i$}.
\END
\eRm

Definitively, the variational approach to the study of invariance properties of equations not only provides conserved quantities associated with the invariance of field equations; when such equations are Euler-Lagrange equations of some (possibly local) Lagrangian we can make a more precise statement about the nature of such conserved quantities: {\em under the condition stressed in the above Remark}, they are determined by  variations of strong Noether currents for the associated local system of Lagrangians. Even in the case of a local presentation of a variational problem, since $\cL_{\Xi}\mathfrak{d}\lam_i = 0$, such a variation of local strong Noether currents is equivalent to a global conserved current. This is an improvement to approaches to mechanics and field theory based on Lie's invariance theory of dynamical systems. 

Thus consequently we state the following main result.
\bTh
The conserved current associated with a generalized symmetry, assumed to be also a symmetry of the variational derivative of the corresponding local inverse problem, is variationally equivalent to the variation of the strong Noether currents for the corresponding local system of Lagrangians. Since the variational Lie derivative of the local system of Lagrangians is a global object, such a variation is variationally equivalent to a global conserved current.
\eTh

\subsection*{Acknowledgements}

Research partially supported by the University of Torino through the Department of Mathematics research project {\em Metodi Geometrici in Fisica Matematica e Applicazioni}; E.W. is also supported by his Department of  Mathematics research grant {\em  Sequenze variazionali e Teoremi di Noether} 2010-2012.

%---------------------------------------------

\end{document}